\documentclass[english,notitlepage,superscriptaddress,nofootinbib,twocolumn,aps,prl]{revtex4-2}
\usepackage[T1]{fontenc}
\usepackage[latin9]{inputenc}
\usepackage{amsmath}
\usepackage{amssymb}
\usepackage{graphicx}
\usepackage{braket}
\usepackage{babel}
\usepackage{hyperref}
\hypersetup{colorlinks=true}
\usepackage{multirow}
\begin{document}
\title{
Relativistic Dips in Entangling Power of Gravity
}

\newcommand{\affone}{University College London, Gower Street, WC1E 6BT London, United Kingdom}
\newcommand{\afftwo}{Physics Institute, University of S\~{a}o Paulo, Rua do Mat\~{a}o, 1371, S\~{a}o Paulo, Brazil}
\newcommand{\afffour}{Van Swinderen Institute, University of Groningen, 9747 AG, The Netherlands}

\author{Marko Toro\v{s}}
\affiliation{\affone}
\author{Martine Schut}
\affiliation{\afffour}
\author{Patrick Andriolo}
\affiliation{\afftwo}
\author{Sougato Bose}
\affiliation{\affone}
\author{Anupam Mazumdar}
\affiliation{\afffour}

\begin{abstract}
{
The salient feature of both classical and quantum gravity is its universal and attractive character. However, less is known about the behaviour and build-up of quantum correlations when quantum systems interact via graviton exchange. In this work, we show that quantum correlations can remain strongly suppressed for certain choices of parameters even when considering two adjacent quantum systems in delocalized states. Using the framework of linearized quantum gravity with post-Newtonian contributions, we find that there are special values of delocalization where gravitationally induced entanglement drops to negligible values, albeit non-vanishing. 
We find a pronounced cancellation point far from the Planck scale, where the system tends towards classicalization. In addition,  we show that quantum correlations begin to reemerge for large and tiny delocalizations due to Heisenberg's uncertainty principle and the universal coupling of gravity to the energy-momentum tensor, forming a valley of gravitational entanglement.
}
\end{abstract}

\maketitle

\section{Introduction}

The coupling of classical gravity to the stress-energy tensor has
been probed in numerous experiments and has withstood the test of
time in all astronomical observations~\citep{will2014confrontation}.
One of its most distinguishing consequences is the universal and attractive
character of the induced gravitational matter-matter interaction.
Such behaviour is manifest by looking at Newton's 1/r potential, but
persists also when including post-Newtonian (PN) corrections depending
on the particle momenta~\citep{iwasaki1971quantum,hiida1972gauge,
Donoghue:1994dn,cristofoli2019post,grignani2020fixing}. 

The same universal and attractive behaviour is also a feature of an effective field theory of quantum gravity~\cite{Donoghue:1994dn}, where the gravitational field is not real-valued
but rather operator-valued~\citep{Bose:2022uxe,Biswas:2022qto,Vinckers:2023grv,Elahi:2023ozf}. 
Such a quantum interaction can generate non-classical correlations between quantum
systems with no classical analogue, making it ideal for testing genuinely
quantum aspects. This observation was critical in conceiving
a protocol to test the quantum nature of gravity using two
massive particles~\citep{Bose:2017nin,marletto2017gravitationally}~\footnote{The results of Ref.~\citep{Bose:2017nin} were already known earlier,
see~\citep{ICTS}.}. This protocol, known as the quantum gravity-induced entanglement
of masses (QGEM), is in this regard akin to Bell's original idea of
testing quantum correlations between two
spatially separated systems~\citep{bell1964einstein,brunner2014bell}.

According to the Local Operations and Classical
Communication (LOCC) principle~\citep{Bennett_1996}, entanglement
can only be generated by ostensibly quantum interactions between the
test particles. 
Hence, only if gravity is a quantum entity will it
generate an entangled state of the two masses~\cite{Bose:2017nin,Marshman:2019sne}. 
Within the context
of an effective field theory of quantum gravity the gravitational interaction is being mediated by the massless spin-2 graviton, see~\citep{gupta1952quantization,gupta1952quantization2,Donoghue:1994dn,Marshman:2019sne,Carney_2019,Christodoulou:2022vte,christodoulou2019possibility,Belenchia:2018szb,Danielson:2021egj,Bose:2022uxe,Biswas:2022qto,Vinckers:2023grv,Elahi:2023ozf}, for a textbook, see~\cite{Scadron:2007qd}.


Theoretical works and feasibility studies about the QGEM proposal
have mainly focused on the static regime where the momenta of the
particles (i.e., the
PN corrections) are neglected in the gravitational interaction. In the static Newtonian limit, the interaction is
in position, and the spatial delocalizations of the quantum states
control entanglement generation. Generally speaking, increasing
the spatial delocalization $\Delta x$ will increase the overall generated
entanglement. Based on the intuition from the static Newtonian limit, decreasing the spatial delocalization $\Delta x$
would suggest that the generated entanglement is bound to
decrease. We will show that this is not the case and that this naive picture breaks down when the PN corrections
are included in the analysis~\citep{poisson2014gravity}. 

In this paper, we will show that the generated entanglement in general increases
for both very small and very large spatial delocalizations $\Delta x$.
This observation is, in a way, simple, and it follows directly from Heisenberg's
uncertainty principle~\cite{busch2007heisenberg} and the universal coupling of gravity to energy~\cite{misner1973gravitation}.
The generated entanglement entropy, quantifying the degree of entanglement,
scales as a function of $\Delta x$ for large spatial delocalizations
and as a function of $\Delta p\propto\hbar/\Delta x$ for small spatial
delocalizations. We will illustrate this result using a toy setup
of two harmonic oscillators interacting gravitationally for the initial
state of the product of ground states. 

There is also an unexpected twist in the story. We find small pockets in the parameter
space of $\Delta x$ -- far from the Planck scale -- where the entanglement is strongly suppressed.
This decrease is pronounced when the dominant 0PN and 1PN terms entering in
the generation of quantum correlations cancel each other's contributions. 
The explanation lies in the opposite sign of the two-mode squeezing parameter induced by the 0PN and 1PN couplings.
We show that this behaviour persists, and can be controlled, by varying the degree of squeezing of the initial states. We conclude by arguing that analogous cancellations of quantum correlations, where the system tends towards classicalization, should appear also in any other quantum mechanical theory of gravity.


\section{Quantum gravitational potential and harmonic oscillators}

We consider the simple 1D toy model of two identical harmonic oscillators,
A and B, characterized by the mass $m$ and angular frequency $\omega_{\text{m}}$,
oscillating along the x-axis. We will assume that the centres of the
two harmonic traps are separated by a distance $d$, and the two
particles interact only gravitationally. Using Gupta's framework of
linearized quantum gravity~\citep{gupta1952quantization,gupta1952quantization2}
and perturbation theory, we can then obtain the gravitational matter-matter
potential up to order 2PN~\citep{Bose:2022uxe} (see also \cite{Biswas:2022qto,Elahi:2023ozf,Vinckers:2023grv}):
\begin{alignat}{1}
\hat{H}_{\text{grav}}  = & -\frac{Gm^{2}}{|\hat{r}_{A}-\hat{r}_{B}|}
-\frac{G(3\hat{p}_{A}^{2}-8\hat{p}_{A}\hat{p}_{B}+3\hat{p}_{B}^{2})}{2c^{2}|\hat{r}_{A}-\hat{r}_{B}|}\nonumber \\
 & +\frac{G(5\hat{p}_{A}^{4}-18\hat{p}_{A}^{2}\hat{p}_{B}^{2}+5p_{B}^{4})}{8c^{4}m^{2}|\hat{r}_{A}-\hat{r}_{B}|}\,,\label{eq:starting}
\end{alignat}
where $c$ ($G$) denotes the speed of light (the gravitational constant).
Here, we will implicitly assume that the momenta are sufficiently small
such that higher order terms, i.e., the terms $\propto\hat{p}_{j}^{n}\hat{p}_{k}^{n'}$
(with $j,k=A,B$ and $n+n'>4$), can be neglected. Higher
order terms would only modify the quantitative results for relativistic
momenta without affecting the features in the regime where the velocities
are small compared to the speed of light (i.e., we neglect terms beyond $\mathcal{O}(c^{-4})$). To keep the expressions short, we also implicitly assume the convention that unsymmetrized expressions (e.g., $\hat{x}\hat{p}$) are to be interpreted in the symmetrized ordering (e.g., $(\hat{x}\hat{p}+\hat{p}\hat{x})/2$).

Post-Newtonian corrections have, of course, been analyzed extensively
in previous works in the center of momentum frame~~\citep{iwasaki1971quantum,hiida1972gauge,cristofoli2019post,grignani2020fixing,Blanchet_2014}.
If we set $\hat{p}\equiv\hat{p}_{A}=-\hat{p}_{B}$, and denote $\hat{r}\equiv\vert{\hat{r}}_{A}-{\hat{r}}_{B}\vert$,
we recover from Eq.~\eqref{eq:starting} the known result in the
literature: 
\begin{equation}
\hat{H}_{\text{grav}}=-\frac{Gm^{2}}{\hat{r}}-7\frac{G\hat{p}^{2}}{c^{2}\hat{r}}-\frac{G\hat{p}^{4}}{c^{4}m^{2}\hat{r}}.\label{eq:classical}
\end{equation}
In other words, Eq.~\eqref{eq:starting} can be seen as the adaptation of Eq.~\eqref{eq:classical}
to the specific case of two quantum harmonic oscillators. 
Generalizations
to more particles, e.g., four harmonic oscillators, forming pairs
of particles and detectors~\citep{gunnink2023gravitational},
modified gravity scenarios such as in the case of a massive graviton~\citep{Elahi:2023ozf},
fat graviton with nonlocal interaction~\citep{Vinckers:2023grv,Vinckers:2024ecg},
or a dilaton-graviton combination~\citep{Chakraborty:2023kel}, can
also be analyzed using similar methods.

We now suppose that the two trap centres are located at $\pm d/2$
and write $\hat{r}_{A}=-d/2+{\hat{x}}_{A}$, and $\hat{r}_{B}=d/2+{\hat{x}}_{B}$.
The operators ${\hat{x}}_{A}$ and $\hat{x}_{B}$ denote small displacements from
the equilibrium position, while the corresponding conjugate momenta
are given by $\hat{p}_{A}$ and $\hat{p}_{B}$, respectively. The
Hamiltonian of the two harmonic oscillators is given by

\begin{equation}
\hat{H}_{\text{matter}}=\frac{\hat{p}_{A}^{2}}{2m}+\frac{\hat{p}_{B}^{2}}{2m}+\frac{m\omega_{\text{m}}^{2}}{2}\hat{x}_{A}^{2}+\frac{m\omega_{\text{m}}^{2}}{2}\hat{x}_{B}^{2},\label{eq:matter}
\end{equation}
where $\omega_\text{m}$ and $m$ denote the harmonic frequency and mass, respectively (assumed for simplicity to be the same for the two harmonic oscillators).
For later convenience, we introduce the mode decompositions 
\begin{alignat}{2}
\hat{x}_{\text{A}} & =\delta x(\hat{a}+\hat{a}^{\dagger}),\,\, \,\, & \hat{x}_{\text{B}} & =\delta x(\hat{b}+\hat{b}^{\dagger}),\label{eq:modesxy}\\
\hat{p}_{\text{A}} & =i\delta p(\hat{a}^{\dagger}-\hat{a}),\,\,\,\, & \hat{p}_{\text{B}} & =i\delta p(\hat{b}^{\dagger}-\hat{b}),\label{eq:modepxpy}
\end{alignat}
where $\hat{a},\hat{b}$ ($\hat{a}^\dagger,\hat{b}^\dagger$) denote the annihilation (creation) operators, and $\delta x=\sqrt{\frac{\hbar}{2m\omega_{\text{m}}}}$, $\delta p=\sqrt{\frac{\hbar m\omega_{\text{m}}}{2}}$ are the
position, momentum zero-point-motions, respectively.

Here, we will be interested in the phenomenology of the gravitational potential
in Eq.~\eqref{eq:starting} up to the quartic order in the operators
$\propto\hat{O}_{i}\hat{O}_{j}\hat{O}_{k}\hat{O}_{l}$ (with $\hat{O}_{i,j,k,l}=\hat{x}_{A},\hat{x}_{B},\hat{p}_{A},\hat{p}_{B}$)
corresponding to small position and momentum fluctuations. Taylor
expanding Eq.~\eqref{eq:starting} around the equilibrium positions
and considering small fluctuations, we find the
following Hamiltonian:
\begin{equation}
\hat{H}_{\text{grav}}=-\frac{Gm^{2}}{d}+\hat{H}_{\text{A}}+\hat{H}_{\text{B}}+\hat{H}_{\text{AB}},\label{eq:split}
\end{equation}
where the first term only produces a global phase, and $\hat{H}_{A}$
and $\hat{H}_{B}$ depend only on the operators of particle
A and B, respectively.
The leading order cross-coupling terms between the two particles
are given by
\begin{equation}
\hat{H}_{\text{AB}}=\frac{Gm^{2}}{d}\left[\frac{2\hat{x}_{\text{A}}\hat{x}_{\text{B}}}{d^2}
 +\frac{4\hat{p}_{\text{A}}\hat{p}_{\text{B}}}{m^2c^{2}}
-\frac{9\hat{p}_{A}^{2}\hat{p}_{B}^{2} }{4m^{4}c^{4}} \right],\label{eq:AB}
\end{equation}
where the 0PN, 1PN, and 2PN contributions appear from left to right, respectively.

While the leading order gravitational
force arises from the terms $\hat{H}_{\text{A}}$ and $\hat{H}_{\text{B}}$
(as these terms contain the uniform gravitational fields affecting the motion of the individual particle), the leading
order contribution for entanglement generation arises from the cross-couplings
in $\hat{H}_{\text{AB}}$~\citep{Balasubramanian:2011wt}. This is already hinting that the
generation of gravitationally induced entanglement might be hiding
some surprises. 

\section{Gravitational entanglement entropy}\label{sec:stat}
In this section we illustrate how the gravitational entanglement entropy depends on the delocalization using simple first order perturbation theory. We assume that the initial state is a product of the two ground states
\begin{equation}
\vert\psi_{\text{i}}\rangle=\vert0\rangle_{\text{A}}\vert0\rangle_{\text{B}},\label{eq:ground}
\end{equation}
which is perturbed by the gravitational interaction $\hat{H}_{\text{grav}}$
in Eq.~\eqref{eq:split}. 
We can always decompose the perturbed state vector in the number basis:
\begin{equation}
\vert\psi_{AB}\rangle=\frac{1}{\sqrt{\mathcal{N}}}\sum_{n,N}C_{nN}\vert n\rangle\vert N\rangle,\label{eq:expansion}
\end{equation}
where $C_{nN}$ denote the coefficients (with $C_{00}=1$), $\mathcal{N}=\sum_{n,N}\vert C_{nN}\vert^{2}$
is the normalization, and $\vert n\rangle$, $\vert N\rangle$ denote
the number states of the two harmonic oscillators.
In particular, the coefficients appearing in Eq.~\eqref{eq:expansion}
can be computed using
\begin{equation}
C_{nN}=\frac{\langle n\vert\langle N\vert\hat{H}_{AB}\vert0\rangle\vert0\rangle}{2E_{0}-E_{n}-E_{N}},\label{eq:coeffs}
\end{equation}
where $E_{0}$ and $E_{n}$, $E_{N}$ denote the energy of the ground
state $\vert0\rangle$ and of the excited states $\vert n\rangle$, $\vert N\rangle$,
respectively (we recall that for a harmonic oscillator, we have $E_j = E_0 + \hbar \omega j$ with $j$ denoting the occupation number).

\begin{figure}
\includegraphics[width=1\columnwidth]{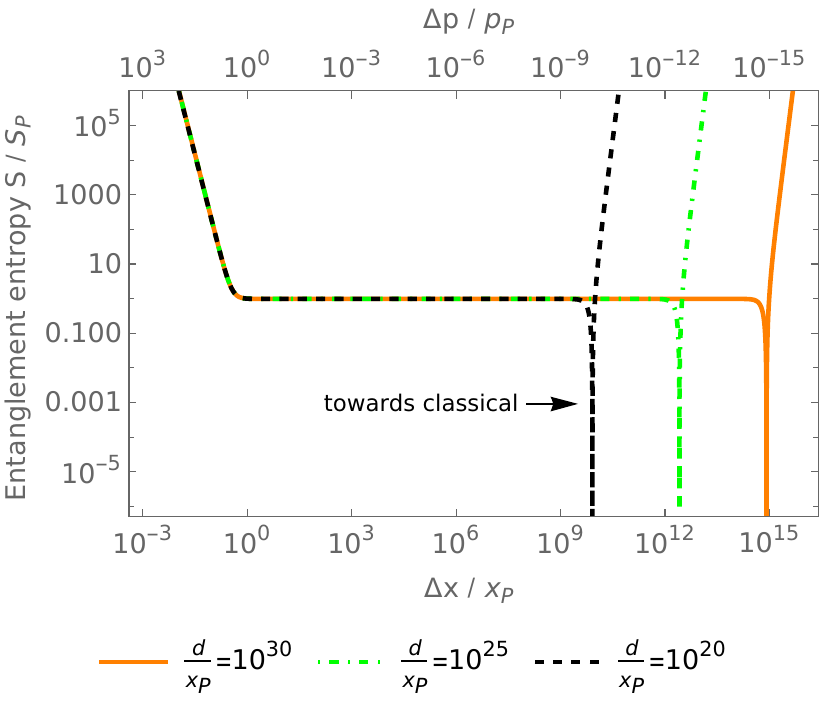}

\caption{\label{fig:simple1}Entanglement entropy $S$ as a function of the
spatial delocalization $\Delta x$ (bottom) or the momentum delocalization
$\Delta p=\hbar/\Delta x$ (top). We have expressed the spatial (momentum) superposition in units of the Planck length (momentum) given by $x_\text{P}\equiv \sqrt{\hbar G/c^3}$ ($p_\text{P}\equiv \hbar/x_\text{P}$).
The behaviour on the right is determined
by the static contribution $\propto\hat{x}_{\text{A}}\hat{x}_{\text{B}}$.
The plateau in the middle is determined by the non-static 1PN contribution
$\propto\hat{p}_{\text{A}}\hat{p}_{\text{B}}$. The dominant static and the non-static couplings cancel each other's contribution to the entanglement entropy at $\sqrt{2}d\omega_{\text{m}}\sim c$, and we observe that the entanglement
entropy $S$ drops to negligible values (the three pronounced dips indicated by the arrow). We note that these dips of quantum correlations indicate a \emph{classicalization} of the system and that they occur far from the Planck length $x_p$.
Finally, as we approach relativistic velocities (corresponding to large momentum superposition sizes on the left side of the figure), we find that higher-order momentum contributions become important. Here we have plotted
the contribution from the term $\propto\hat{p}_{\text{A}}^{2}\hat{p}_{\text{B}}^{2}$
arising at order $\mathcal{O}(1/c^{4})$. We plot the curves for different values of the trap distance $d$. Note that the qualitative behaviour remains the same in all three cases, with the location of the dip shifting to the left (right) for smaller (larger) values of $d$ as expected  (i.e., larger distances weaken the position couplings, hence requiring larger spatial superposition sizes).  The entanglement entropy is expressed in units of the entanglement entropy when we set the spatial superposition to that of the Planck length, i.e.,
$S_p=S(\Delta x = x_\text{P}) = -(\frac{Gm}{c^{2}d})^{2}\log ((\frac{Gm}{c^{2}d})^{2})$ which is achieved on the plateau. }
\end{figure}
We will quantify the degree of entanglement using the von Neumann entanglement entropy given
by $S=-\text{tr}[\rho_{A}\text{ln}\rho_{A}]$, where $\rho_{A}=\text{tr}_{B}[\rho_{AB}]$ is the reduced density
matrix of subsystem $A$, and $\rho_{AB}=\vert\psi_{AB}\rangle\langle\psi_{AB}\vert$
is the total density matrix of the system (see Appendix~\ref{AppendixA} for a short review on the entanglement entropy). Using Eqs.~\eqref{eq:AB}-\eqref{eq:coeffs} find a simple formula for the steady-state entanglement entropy:
\begin{alignat}{1}
S\approx & -\left(\frac{Gm}{c^{2}d}-\frac{Gm}{2d^{3}\text{\ensuremath{\omega_{\text{m}}^{2}}}}\right)^{2}\log\left(\left(\frac{Gm}{c^{2}d}-\frac{Gm}{2d^{3}\ensuremath{\omega_{\text{m}}^{2}}}\right)^{2}\right)\nonumber \\
 & -\frac{81G^{2}\ensuremath{\omega_{\text{m}}^{2}}\hbar^{2}}{1024c^{8}d^{2}}\log\left(\frac{81G^{2}\ensuremath{\omega_{\text{m}}^{2}}\hbar^{2}}{1024c^{8}d^{2}}\right),\label{eq:simple1}
\end{alignat}
which is plotted in Fig.~\ref{fig:simple1}  (see Appendix~\ref{AppendixB} for the detailed derivation). The first line of Eq.~\eqref{eq:simple1} captures the right-most
part of the plot (from the 0PN $\hat{x}_A \hat{x}_B$ coupling) as well as the plateau (from the 1PN $\hat{p}_A \hat{p}_B$ coupling). The right-most part of the plot in Fig.~\ref{fig:simple1} is captured
by the 2PN contribution in the second line of Eq.~\eqref{eq:simple1}
arising from the coupling $\propto\hat{p}_{A}^{2}\hat{p}_{B}^{2}$.

We find that the entanglement entropy grows for very large and very small positions
delocalizations $\Delta x$ given by the zero-point-motion. For large
$\Delta x$ the entanglement entropy grows as a consequence of the
0PN static position couplings arising from the familiar 1/r potential,
while for small $\Delta x$ the momentum delocalization $\Delta p\sim\hbar/\Delta x$
becomes large, and the post-Newtonian momentum couplings start increasing
the entanglement entropy. In other words, the landscape of entanglement
entropy as a function of delocalizations $\Delta x$ or $\Delta p$ forms a valley. This can be seen
as a consequence of Heisenberg's minimum uncertainty relation $\Delta p\Delta x=\hbar/2$
and the universal coupling of gravity to all forms of energy. 

In addition,  Fig.~\ref{fig:simple1} reveals unexpected dips in entanglement entropy, which appear far the Planck length. 
These dips arise because of the cancellation between
the 0PN and 1PN terms in the first line of Eq.~\eqref{eq:simple1} corresponding to the couplings
$\propto\hat{x}_{A}\hat{x}_{B}$ and $\propto\hat{p}_{A}\hat{p}_{B}$,
respectively. The first line vanishes when the product of the harmonic
frequency, $\omega_\text{m}$, and of the distance between the two traps, $d$,
becomes comparable to the speed of light, i.e., $\sqrt{2}\, \omega_{\text{m}} d\ensuremath{}=c$.
In this case, the position and momentum delocalizations are given by
$\Delta x=\sqrt{\frac{\hbar d}{mc}}/\sqrt[4]{2}$ and 
$\Delta p=\sqrt{\frac{\hbar mc}{d}}/\sqrt[4]{2}$,
respectively. At this point, the two adimensional parameters governing
the 0PN and 1PN match, i.e., $\Delta x/d=\sqrt{2}\Delta p/(mc)$.

The adimensional parameters however only explain that the 0PN and 1PN contributions are
of equal magnitude without revealing the origin of the cancellation. As we will see in the next section, this cancellation arises from the opposite sign of the two-mode squeezing (TMS) character of the position and
momentum couplings. 

\section{Dips of gravitational entanglement}\label{time-dependent}

\begin{figure}
\includegraphics[width=0.89\columnwidth]{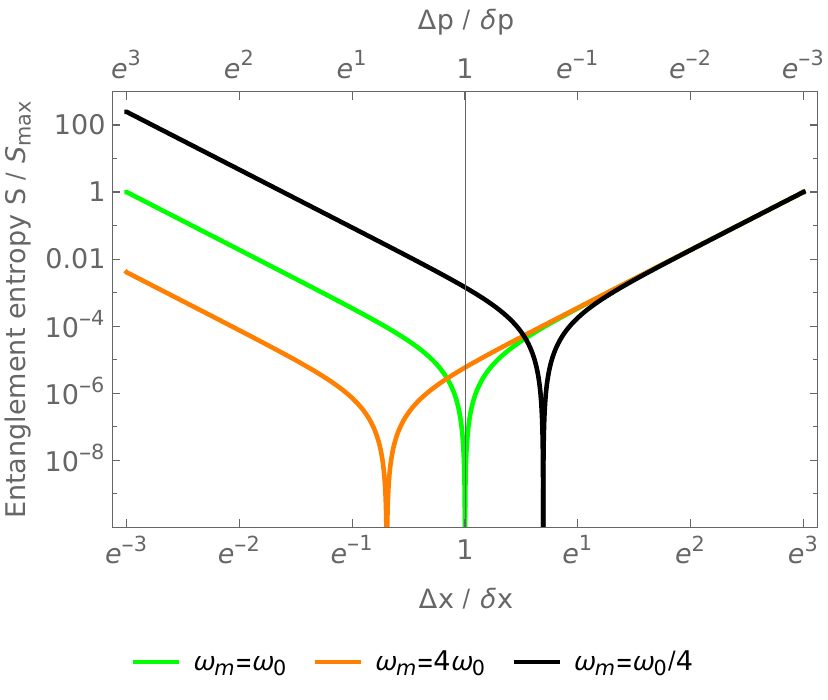}
\caption{The maximum entanglement value achieved during the time evolution starting with a product of two single-mode-squeezed-vacuum (SMSV) states as a function of
the spatial delocalization $\Delta x=\delta xe^{-r}$ (bottom axis)
and momentum delocalization $\Delta p=\delta pe^{r}$ (top axis).
By setting $\omega_{\text{m}}=\omega_{0}\equiv c/(\sqrt{2}d)$
we find the case of equal couplings $g_{x}=g_{p}=g_{0}\equiv\sqrt{2}Gm/(c^{2}d)$
(green line). We also consider the case $\omega_{\text{m}}=4\omega_{0}$
producing the couplings $4g_{x}=g_{p}/4=g_{0}$ (orange line), and
the case $\omega_{\text{m}}=\omega_{0}/4$ producing the coupling
$g_{x}/4=4g_{p}=g_{0}$ (black line). The entanglement entropy is normalized to the maximum value of the green line to ease the comparison. We note that both the horizontal location and the depth can be changed by tuning the couplings $g_x$, $g_p$ using the mechanical frequency $\omega_\text{m}$. At the location of the dips the system tends towards classicalization as there is a strong suppression of quantum correlations.
\label{fig:timecase}}
\end{figure}
In this section we look closer at the unexpected dips of gravitational entanglement uncovered in the previous section. We restrict the analysis only to the leading order terms
identified in the previous section as its origin, i.e.,
\begin{equation}
\hat{H}_{\text{AB}}=\frac{2Gm^{2}}{d^{3}}\hat{x}_{\text{A}}\hat{x}_{\text{B}}+\frac{4G}{c^{2}d}\hat{p}_{\text{A}}\hat{p}_{\text{B}}.\label{eq:quadratic}
\end{equation}
Using Eqs.~\eqref{eq:modesxy} and  \eqref{eq:modepxpy} in Eq. \eqref{eq:quadratic}
we then find
\begin{equation}
\hat{H}_{\text{AB}}=\hbar g_{-}(\underbrace{ab+a^{\dagger}b^{\dagger}}_{\text{TMS}})+\hbar g_{+}(\underbrace{ab^{\dagger}+a^{\dagger}b}_{\text{BS}}),\label{eq:HamiltonianModes}
\end{equation}   
where the coupling rates are 
\begin{alignat}{1}
g_{-}=g_{x}-g_{p} & ,\qquad g_{+}=g_{x}+g_{p},\\
g_{x}=\frac{Gm}{d^{3}\omega_{\text{m}}} & ,\qquad g_{p}=\frac{2Gm\omega_{\text{m}}}{c^{2}d}, \label{eq:couplings}
\end{alignat}
and TMS (BS) labels the two-mode squeezing (beam-splitter) contribution (for an introduction on quantum optics transformations see for example~\cite{gerry2023introductory,leonhardt2010essential}).

For an initial product of ground states $\vert 0\rangle \vert 0\rangle$ the BS part of the Hamiltonian has no effect, 
while the TMS-induced change depends on the PN order: at 0PN we have the coupling $+g_{x}$, while at 1PN we have the coupling $-g_{p}$. The explanation for the dips in Fig.~\ref{fig:simple1} thus lies in the opposite sign of the TMS transformation arising from the position and momentum couplings of gravity. When $g_x=g_p$ the 0PN and 1PN contributions to TMS cancel, i.e., $g_{-}=0$, and we find a dip.

The observed cancellation point persists even if we consider the initial state to
be the product of two single-mode-squeezed-vacuum (SMSV) states which have enhanced
position or momentum delocalizations given by 
\begin{equation}
\Delta x=\delta xe^{-r},\qquad\Delta p=\delta pe^{r},
\end{equation}
respectively (with $\delta x$ and $\delta p$ denoting the zero-point-motions defined below Eq.~\eqref{eq:modepxpy}). Such states are
not energy eigenstates, and hence, we have to take into account their
time-evolution in the harmonic traps (see derivation in Appendix~\ref{AppendixC}).
In Fig.~\ref{fig:timecase} we show how the maximum achieved entanglement entropy changes by varying the degree of squeezing $r$ of the initial state. We note that by tuning the mechanical frequency $\omega_\text{m}$ (and hence the couplings $g_x, g_p$ in Eq.~\eqref{eq:couplings}) we can also shift to profile horizontally and increase its depth. The location of the dip is now determined by the condition $g_x e^{-2r}=g_p e^{2r}$, which depends on the squeezing parameter $r$ as expected.

\section{Discussion}
In this work, we have discussed entanglement generation within the
context of a linearized quantum gravity with post-Newtonian (PN) momentum
contributions. We have considered a simple toy model of two nearby
harmonic oscillators and computed the entanglement entropy arising
from their gravitational interaction. We uncovered a surprising interplay
between the Heisenberg's uncertainty relation, the squeezing character of
gravity, and the generation of non-classical correlations when 1PN
and 2PN contributions are taken into account. 


We found that  the entanglement
entropy drops to negligible values when the 0PN contribution and the
1PN are of the same magnitude, i.e., when $\Delta x/d\sim\Delta p/(mc)$,
where $m$ is the mass of each harmonic oscillator, and $d$ denotes
the distance between the two trap centres.  Although the classical gravitational force is universally
attractive, the generation of gravitationally induced entanglement
can be suppressed to negligible values for specific states. 
As this will remain true within any theory that recovers the predictions of perturbative quantum gravity, such states could. provide a method to distinguish perturbative quantum gravity from other classes of gravitational theories, such as a scalar-tensor theory with different PN contributions. While this work does not answer questions of experimental feasibility, it nonetheless uncovers experimentally defining features of infrared quantum gravity. The entanglement dips provide a distinct signature of classicalization, showing how quantum correlations can become suppressed in a fully quantum mechanical framework.

\begin{acknowledgments}
   MT would like to acknowledge funding from ST/W006227/1.
   MS is supported by the Fundamentals of the Universe research program at the University of Groningen.
   PA acknowledges funding from Brazilian Coordination for the Improvement of Higher Education Personnel (CAPES), 
   Erasmus+ Programme and the Strategic Partnership Framework at University of Groningen.
   SB thanks EPSRC grants EP/R029075/1, EP/X009467/1, and ST/W006227/1. 
   SB and AM's research is supported by the Sloan, and Betty and Moore foundations. 
\end{acknowledgments}

\bibliographystyle{unsrt}
\bibliography{heisenberg}

\appendix

\section{Brief overview of entanglement entropy formulae} \label{AppendixA}

The von Neumann entanglement entropy is given
by

\begin{equation}
S=-\text{tr}[\rho_{A}\text{ln}\rho_{A}],\label{eq:ent}
\end{equation}
where $\rho_{A}=\text{tr}_{B}[\rho_{AB}]$ is the reduced density
matrix of subsystem $A$, $\rho_{AB}=\vert\psi_{AB}\rangle\langle\psi_{AB}\vert$
is the total density matrix of the system. The same form of the expression
would be obtained by tracing over system A by formally exchanging the subsystems
A and B, i.e., $S=-\text{tr}[\rho_{B}\text{ln}\rho_{B}]$ and $\rho_{B}=\text{tr}_{A}[\rho_{AB}]$.
The general expression in Eq.~\eqref{eq:ent} can be however rewritten in
a more convenient way for the two cases we analyse in Secs.~\ref{sec:stat} and \ref{time-dependent}.

In Sec.~\ref{sec:stat}, we will suppose that the initial state is the product of the ground states of two harmonic oscillators, i.e., $\vert 0 \rangle \vert 0\rangle$ and use time-independent perturbation theory. Using the Schmidt decomposition of the state in Eq.~\eqref{eq:expansion}
it can be shown that the entanglement entropy from Eq.~\eqref{eq:ent} reduces to 
\begin{equation}
S=-\sum_{j}\vert\alpha_j\vert^2\text{log}\left[\vert\alpha_j\vert^2 \right],\label{eq:entnumber}
\end{equation}
where $\alpha_j$ are the Schmidt coefficients. The Schmidt decomposition can, in general, be performed using the linear algebra technique of singular value decomposition (SVD)~\cite{nielsen2010quantum}. However, we will be primarily interested in the states of the form $\sum_j \alpha_j \vert j\rangle\vert j\rangle$ (with $\vert j\rangle$ denoting the number states), where one can readily read-off the Schmidt coefficients $\alpha_j$. 

In Sec.~\ref{time-dependent}, we will be interested in Gaussian states and their time-evolution. The formula for the entanglement entropy simplifies to~\cite{audenaert2002entanglement,PhysRevA.70.022318,PhysRevA.71.032349,RevModPhys.84.621,RevModPhys.77.513}:
\begin{equation}
S=-f\text{ln}f+(1+f)(1+\text{ln}f),\label{eq:st}
\end{equation}
where $f(t)$ is the symplectic eigenvalue of the single-mode covariance
matrix of subsystem A given by 
\begin{equation}
f(t)=\frac{1}{\hbar}\sqrt{4\langle\hat{x}_{A}^{2}\rangle\langle\hat{p}_{A}^{2}\rangle-\langle\hat{x}_{A}\hat{p}_{A}+\hat{p}_{A}\hat{x}_{A}\rangle^{2}}-\frac{1}{2}.\label{eq:ft}
\end{equation}
Also, the formula in Eq.~\eqref{eq:st} remains valid if we formally replace the quantities related to subsystem A
with the ones for subsystem B in Eq.~\eqref{eq:ft} (i.e., we compute the symplectic eigenvalue of the single mode covariance matrix of subsystem B).

\section{List of expansion coefficients up to quartic order in the operators and up to 2PN} \label{AppendixB}

\begin{table*}
\begin{tabular}{|c|c|c|c|c|c|c|c|c|}
\hline 
\multirow{2}{*}{order} & \multirow{2}{*}{coefficient} & \multirow{2}{*}{coupling} & \multicolumn{6}{c|}{Non-zero coefficients $C_{nN}$ from first-order perturbation theory}\tabularnewline
\cline{4-9} \cline{5-9} \cline{6-9} \cline{7-9} \cline{8-9} \cline{9-9} 
 &  &  & $C_{11}$ & $C_{21}$ & $C_{12}$ & $C_{31}$ & $C_{13}$ & $C_{22}$\tabularnewline
\hline 
\hline 
\multirow{6}{*}{0PN} & $\frac{2Gm^{2}}{d^{3}}$ & $\mathbf{\hat{x}_{\text{A}}\hat{x}_{\text{B}}}$ & \textbf{$\boldsymbol{-\frac{Gm}{2d^{3}\omega_{\text{m}}^{2}}}$} &  &  &  &  & \tabularnewline
\cline{2-9} \cline{3-9} \cline{4-9} \cline{5-9} \cline{6-9} \cline{7-9} \cline{8-9} \cline{9-9} 
 & $\frac{3Gm^{2}}{d^{4}}$ & $\hat{x}_{\text{A}}^{2}\hat{x}_{\text{B}}$ &  & $-\frac{Gm\delta x}{\sqrt{2}d^{4}\omega_{\text{m}}^{2}}$ &  &  &  & \tabularnewline
\cline{2-9} \cline{3-9} \cline{4-9} \cline{5-9} \cline{6-9} \cline{7-9} \cline{8-9} \cline{9-9} 
 & $-\frac{3Gm^{2}}{d^{4}}$ & $\hat{x}_{\text{A}}\hat{x}_{\text{B}}^{2}$ &  &  & $\frac{Gm\delta x}{\sqrt{2}d^{4}\omega_{\text{m}}^{2}}$ &  &  & \tabularnewline
\cline{2-9} \cline{3-9} \cline{4-9} \cline{5-9} \cline{6-9} \cline{7-9} \cline{8-9} \cline{9-9} 
 & $\frac{4Gm^{2}}{d^{5}}$ & $\hat{x}_{\text{A}}^{3}\hat{x}_{\text{B}}$ & $-\frac{3G\hbar}{2d^{5}\omega_{\text{m}}^{3}}$ &  &  & $-\frac{\sqrt{6}G\hbar}{4d^{5}\omega_{\text{m}}^{3}}$ &  & \tabularnewline
\cline{2-9} \cline{3-9} \cline{4-9} \cline{5-9} \cline{6-9} \cline{7-9} \cline{8-9} \cline{9-9} 
 & $-\frac{6Gm^{2}}{d^{5}}$ & $\hat{x}_{\text{A}}^{2}\hat{x}_{\text{B}}^{2}$ &  &  &  &  &  & $\frac{3G\hbar}{4d^{5}\omega_{\text{m}}^{3}}$\tabularnewline
\cline{2-9} \cline{3-9} \cline{4-9} \cline{5-9} \cline{6-9} \cline{7-9} \cline{8-9} \cline{9-9} 
 & $\frac{4Gm^{2}}{d^{5}}$ & $\hat{x}_{\text{A}}\hat{x}_{\text{B}}^{3}$ & $-\frac{3G\hbar}{2d^{5}\omega_{\text{m}}^{3}}$ &  &  &  & $-\frac{\sqrt{6}G\hbar}{4d^{5}\omega_{\text{m}}^{3}}$ & \tabularnewline
\hline 
\multirow{12}{*}{1PN} & $\frac{4G}{c^{2}d}$ & $\mathbf{\hat{p}_{\text{A}}\hat{p}_{\text{B}}}$ & $\boldsymbol{\frac{Gm}{c^{2}d}}$ &  &  &  &  & \tabularnewline
\cline{2-9} \cline{3-9} \cline{4-9} \cline{5-9} \cline{6-9} \cline{7-9} \cline{8-9} \cline{9-9} 
 & $\frac{4G}{c^{2}d^{2}}$ & $\hat{p}_{\text{A}}\hat{p}_{\text{B}}\hat{x}_{\text{A}}$ &  & $\frac{2\sqrt{2}Gm\delta x}{3c^{2}d^{2}}$ &  &  &  & \tabularnewline
\cline{2-9} \cline{3-9} \cline{4-9} \cline{5-9} \cline{6-9} \cline{7-9} \cline{8-9} \cline{9-9} 
 & $-\frac{4G}{c^{2}d^{2}}$ & $\hat{p}_{\text{A}}\hat{p}_{\text{B}}\hat{x}_{\text{B}}$ &  &  & $-\frac{2\sqrt{2}Gm\delta x}{3c^{2}d^{2}}$ &  &  & \tabularnewline
\cline{2-9} \cline{3-9} \cline{4-9} \cline{5-9} \cline{6-9} \cline{7-9} \cline{8-9} \cline{9-9} 
 & $\frac{4G}{c^{2}d^{3}}$ & $\hat{p}_{\text{A}}\hat{p}_{\text{B}}\hat{x}_{\text{A}}^{2}$ & $\frac{G\hbar}{2c^{2}d^{3}\omega_{\text{m}}}$ &  &  & $\frac{\sqrt{6}G\hbar}{4c^{2}d^{3}\omega_{\text{m}}}$ &  & \tabularnewline
\cline{2-9} \cline{3-9} \cline{4-9} \cline{5-9} \cline{6-9} \cline{7-9} \cline{8-9} \cline{9-9} 
 & $-\frac{8G}{c^{2}d^{3}}$ & $\hat{p}_{\text{A}}\hat{p}_{\text{B}}\hat{x}_{\text{A}}\hat{x}_{B}$ &  &  &  &  &  & $-\frac{G\hbar}{c^{2}d^{3}\omega_{\text{m}}}$\tabularnewline
\cline{2-9} \cline{3-9} \cline{4-9} \cline{5-9} \cline{6-9} \cline{7-9} \cline{8-9} \cline{9-9} 
 & $\frac{4G}{c^{2}d^{3}}$ & $\hat{p}_{\text{A}}\hat{p}_{\text{B}}\hat{x}_{B}^{2}$ & $\frac{G\hbar}{2c^{2}d^{3}\omega_{\text{m}}}$ &  &  &  & $\frac{\sqrt{6}G\hbar}{4c^{2}d^{3}\omega_{\text{m}}}$ & \tabularnewline
\cline{2-9} \cline{3-9} \cline{4-9} \cline{5-9} \cline{6-9} \cline{7-9} \cline{8-9} \cline{9-9} 
 & $\frac{3G}{2c^{2}d^{2}}$ & $\hat{p}_{\text{A}}^{2}\hat{x}_{\text{B}}$ &  & $\frac{\sqrt{2}Gm\delta x}{4c^{2}d^{2}}$ &  &  &  & \tabularnewline
\cline{2-9} \cline{3-9} \cline{4-9} \cline{5-9} \cline{6-9} \cline{7-9} \cline{8-9} \cline{9-9} 
 & $-\frac{3G}{2c^{2}d^{3}}$ & $\hat{p}_{\text{A}}^{2}\hat{x}_{\text{B}}^{2}$ &  &  &  &  &  & $-\frac{3G\hbar}{16c^{2}d^{3}\omega_{\text{m}}}$\tabularnewline
\cline{2-9} \cline{3-9} \cline{4-9} \cline{5-9} \cline{6-9} \cline{7-9} \cline{8-9} \cline{9-9} 
 & $\frac{3G}{c^{2}d^{3}}$ & $\hat{p}_{\text{A}}^{2}\hat{x}_{\text{A}}\hat{x}_{\text{B}}$ & $-\frac{3G\hbar}{8c^{2}d^{3}\omega_{\text{m}}}$ &  &  & $\frac{3\sqrt{6}G\hbar}{16c^{2}d^{2}\omega_{\text{m}}}$ &  & \tabularnewline
\cline{2-9} \cline{3-9} \cline{4-9} \cline{5-9} \cline{6-9} \cline{7-9} \cline{8-9} \cline{9-9} 
 & $\frac{3G}{2c^{2}d^{2}}$ & $\hat{p}_{\text{B}}^{2}\hat{x}_{\text{A}}$ &  &  & $\frac{\sqrt{2}Gm\delta x}{4c^{2}d^{2}}$ &  &  & \tabularnewline
\cline{2-9} \cline{3-9} \cline{4-9} \cline{5-9} \cline{6-9} \cline{7-9} \cline{8-9} \cline{9-9} 
 & $-\frac{3G}{2c^{2}d^{2}}$ & $\hat{p}_{\text{B}}^{2}\hat{x}_{\text{A}}^{2}$ &  &  &  &  &  & $-\frac{3G\hbar}{16c^{2}d^{3}\omega_{\text{m}}}$\tabularnewline
\cline{2-9} \cline{3-9} \cline{4-9} \cline{5-9} \cline{6-9} \cline{7-9} \cline{8-9} \cline{9-9} 
 & $\frac{3G}{c^{2}d^{3}}$ & $\hat{p}_{\text{B}}^{2}\hat{x}_{\text{B}}\hat{x}_{\text{A}}$ & $-\frac{3G\hbar}{8c^{2}d^{3}\omega_{\text{m}}}$ &  &  &  & $\frac{3\sqrt{6}G\hbar}{16c^{2}d^{2}\omega_{\text{m}}}$ & \tabularnewline
\hline 
2PN & $-\frac{9G}{4c^{4}m^{2}d}$ & $\mathbf{\hat{p}_{A}^{2}\hat{p}_{B}^{2}}$ &  &  &  &  &  & 
\textbf{$\boldsymbol{\frac{9G\pmb{\hbar}\omega_{\text{m}}}{32c^{4}d}}$}\tabularnewline
\hline 
\end{tabular}\caption{List of all gravitational couplings up to quartic order in operators
and up to order 2PN (first three columns). Using first-order perturbation
theory in Eq.~\eqref{eq:coeffs} we obtain the 25 non-zero coefficients $C_{nN}$ in
columns four to nine (the coefficients $C_{0N}$ and $C_{n0}$ will not contribute to the entanglement at first order in perturbation theory and are left out of the table~\citep{Balasubramanian:2011wt}). The \emph{local dip feature} in the generation
of entanglement occurs as a result of the cancellation arising from
the terms $\propto\hat{x}_{\text{A}}\hat{x}_{\text{B}}$ and $\propto\hat{p}_{\text{A}}\hat{p}_{\text{B}}$.
As discussed in the main text, this is a result of the different signs of the
induced two-mode-squeezing (TMS) transformation from the leading order 0PN and 1PN
terms with the hint in the different sign of the corresponding $C_{11}$
coefficients. We note that additional cancellations occur in every column; we see that the $C_{nN}$ coefficients in all columns have both positive and negative values.
The\emph{ broad valley feature} can be understood directly
from the couplings. On the one hand, for large-position delocalization
$\Delta x$ the $0$PN term $\propto\hat{x}_{\text{A}}\hat{x}_{\text{B}}$ generates a rapid increase of
the entanglement entropy with increasing $\Delta x$, and, on the other hand, for
large momentum delocalization $\Delta p$ (i.e., tiny spatial delocalization)
the term $\propto\hat{p}_{A}^{2}\hat{p}_{B}^{2}$ also produces a fast
growth of the entanglement entropy with increasing $\Delta p$. In other words, entanglement entropy
as function of position delocalization is loosely speaking \textquotedblleft U\textquotedblright{}
shaped, forming a valley of entanglement. 
\label{table1}}
\end{table*}

From Eq.~\eqref{eq:starting} we find that the cross-coupling terms between the two particles are given by
\begin{alignat}{1}
\Delta & \hat{H}_{\text{AB}}=\frac{Gm^{2}}{d^{3}}\bigg(2\hat{x}_{\text{A}}\hat{x}_{\text{B}}+3\frac{(\hat{x}_{\text{A}}^{2}\hat{x}_{\text{B}}-\hat{x}_{\text{A}}\hat{x}_{\text{B}}^{2})}{d}\nonumber \\
 & \quad+\frac{4\hat{x}_{\text{A}}^{3}\hat{x}_{\text{B}}-6\hat{x}_{\text{A}}^{2}\hat{x}_{\text{B}}^{2}+4\hat{x}_{\text{A}}\hat{x}_{\text{B}}^{3}}{d^{2}}\bigg)\nonumber \\
 & +\frac{G}{c^{2}d}\bigg(4\hat{p}_{\text{A}}\hat{p}_{\text{B}}(1+\frac{\hat{x}_{\text{A}}-\hat{x}_{\text{B}}}{d}+\frac{(\hat{x}_{\text{A}}-\hat{x}_{\text{B}})^{2}}{d^{2}})\nonumber \\
 & \quad+\frac{3\hat{p}_{\text{A}}^{2}}{2}(\frac{\hat{x}_{\text{B}}}{d}+\frac{2\hat{x}_{\text{A}}\hat{x}_{\text{B}}-\hat{x}_{\text{B}}^{2}}{d^{2}})\nonumber \\
 & \quad+\frac{3\hat{p}_{\text{B}}^{2}}{2}(\frac{\hat{x}_{A}}{d}+\frac{2\hat{x}_{\text{A}}\hat{x}_{\text{B}}-\hat{x}_{\text{A}}^{2}}{d^{2}})\bigg)-\frac{9G}{4c^{4}m^{2}d}\hat{p}_{A}^{2}\hat{p}_{B}^{2},\label{eq:ABfull}
\end{alignat}
where the first two lines contain the static limit 0PN contribution,
lines three to five contain the 1PN contribution and the last term
corresponds to the 2PN contribution. While the leading order gravitational
force arises from the terms $\hat{H}_{\text{A}}$ and $\hat{H}_{\text{B}}$
(as these terms contain the uniform gravitational fields affecting the motion of the individual particle), the leading
order contribution for entanglement generation arises from the cross-couplings
in $\hat{H}_{\text{AB}}$.

In Table.~\ref{table1}, we have listed the couplings
from Eq.~\eqref{eq:AB} and applied Eq.~\eqref{eq:coeffs} to obtain
the non-zero expansion coefficient $C_{nN}$ for $n,N>0$ (the coefficients $C_{0N}$ and $C_{n0}$ will not contribute to the entanglement at first order in perturbation theory and are left out of the computation).
The resulting entanglement
entropy computed using Eq.~\eqref{eq:entnumber} as a function of
position delocalization $\Delta x\equiv\delta x$ (bottom axis) and
momentum delocalization $\Delta p\equiv\delta p=\hbar/\Delta x$ (top
axis) is plotted in Fig.~\ref{fig:simple1}.

We can, however, capture both qualitatively and quantitatively the
entanglement entropy shown in Fig.~\ref{fig:simple1} by considering
only three couplings from Table.~\ref{table1}. In the right-most
part of the figure, the momentum couplings (1PN and 2PN effects) are
negligible as the momentum delocalization $\Delta p$ is tiny compared
to $mc$, and hence dominated by the position 0PN couplings. Furthermore,
if the size of the delocalization $\Delta x$ is also small compared
to the distance between the traps $d$, then we can further neglect
the cubic and quartic couplings, leaving us with the coupling $\propto\hat{x}_{A}\hat{x}_{B}$.
The intermediate plateau is captured by the 1PN coupling $\propto\hat{p}_{A}\hat{p}_{B}$
as the ratio $\Delta x/d$ becomes tiny, and $\Delta p$ becomes non-negligible
compared to $mc$. Finally, the left-most part of the figure is dominated
by the 2PN coupling $\propto\hat{p}_{A}^{2}\hat{p}_{B}^{2}$, and
eventually by higher order PN corrections as we would further increase
$\Delta p$. By making such simplifications, the perturbed state can
be written as
\begin{equation}
\vert\psi_{\text{AB}}\rangle\approx\frac{1}{\mathcal{N}}[\vert0\rangle\vert0\rangle-C_{11}\vert1\rangle\vert1\rangle-C_{22}\vert2\rangle\vert2\rangle],\label{eq:finalNewton}
\end{equation}
where $\vert0\rangle$, $\vert1\rangle$, $\vert2\rangle$ denote
the number states, and $\mathcal{N}$ denotes the overall normalization.
To compute the entanglement entropy in Eq.~\eqref{eq:simple1}, we can now readily use Eq.~\eqref{eq:entnumber},
where we can make the further approximation $\mathcal{N}\approx1$
as we have $C_{00}\approx 1$ and $C_{11},C_{22}\ll1$ (while the terms $C_{11}^{2},C_{22}^{2}$
appearing in $\mathcal{N}$ would only contribute higher order corrections).

Let us briefly comment how to see the entanglement dip from Table.~\ref{table1}. We first recall that the TMS operator can be written in the form  $\hat{S(\xi)}=\exp (-i (\xi^* \hat{a}\hat{b} + \xi\hat{a}^{\dagger}\hat{b}^{\dagger}))$, where the TMS generator is 
$\propto \xi^* \hat{a}\hat{b} +\xi\hat{a}^{\dagger}\hat{b}^{\dagger}$. From Eq.~\eqref{eq:coeffs} we immediately find 
\begin{equation}
C_{nN}\propto
\langle n\vert\langle N\vert\hat{H}_{AB}\vert0\rangle\vert0\rangle 
\propto \pm\langle n\vert\langle N\vert\hat{a}^{\dagger}\hat{b}^{\dagger}+\hat{a}\hat{b}\vert0\rangle\vert0\rangle,
\label{eq:coeffs-1}
\end{equation}
with the plus (minus) sign corresponding to the 0PN coupling $\hat{x}_{A}\hat{x}_{B}$
(1PN coupling $\hat{p}_{A}\hat{p}_{B}$). In other words, the 0PN
position coupling would like to squeeze with TMS parameter $\xi=+1$
while the 1PN momentum coupling would like to squeeze in the opposite
direction with TMS parameter $\xi=-1$. The 0PN and 1PN two-mode squeezing
contributions cancel when $\Delta x/d=\sqrt{2}\Delta p/(mc)$.
Summing the two contributions, we find a total squeezing parameter $\xi=0$ and a suppression of the gravitationally
induced entanglement. The reason for the entanglement suppression
thus lies in the opposite sign of the two-mode squeezing (TMS) parameter at the leading
order 0PN and 1PN gravitational interaction.


\section{Derivation of the time-dependent entanglement entropy} \label{AppendixC}

\begin{figure}
\includegraphics[width=1\columnwidth]{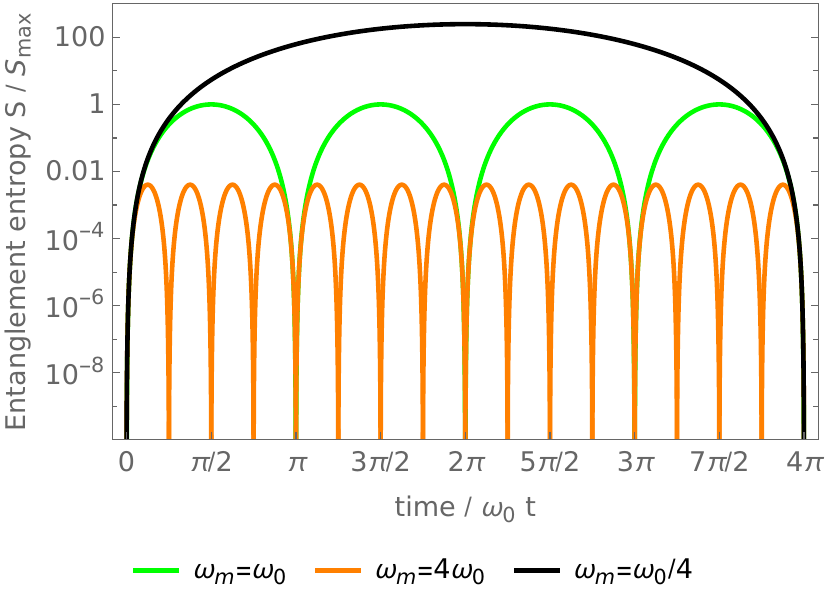}\hspace{1.5cm}
\caption{Entanglement entropy $S$ as a function of time $t$
for different values of the couplings $g_x$, $g_p$ defined in Eq.~\eqref{eq:couplings} (which scale as a function of the harmonic frequency $\omega_\text{m}$).
The SMSV squeezing parameter is set to $r=-3$ corresponding to initial position squeezing  $\Delta x=\delta xe^{-r}$ and momentum delocalization $\Delta p=\delta pe^{r}$. By setting $\omega_{\text{m}}=\omega_{0}\equiv c/(\sqrt{2}d)$
we find the case of equal couplings $g_{x}=g_{p}=g_{0}\equiv\sqrt{2}Gm/(c^{2}d)$
(green line). We also consider the case $\omega_{\text{m}}=4\omega_{0}$
producing the couplings $4g_{x}=g_{p}/4=g_{0}$ (orange line), and
the case $\omega_{\text{m}}=\omega_{0}/4$ producing the coupling
$g_{x}/4=4g_{p}=g_{0}$ (black line). In all cases, the maximum entanglement
is generated at $\omega_{\text{m}}t=\pi/2$. The entanglement entropy is normalized to the maximum value $S_\text{max}$ of the green curve to ease the comparison with Fig.~\ref{fig:timecase}. \label{fig:timecaseAppendix}}
\end{figure}

Here, we further explore the dip in entanglement generation by considering the initial state to
be the product of two single-mode-squeezed-vacuum (SMSV) states:
\begin{equation}
\vert\psi_{\text{i}}\rangle=\vert r\rangle_{\text{A}}\vert r\rangle_{\text{B}},\label{eq:squeezed}
\end{equation}
where $r\in\text{Re}$ is the SMSV squeezing parameter. The single
mode squeezed state is given by
\begin{equation}
\vert r\rangle=\frac{1}{\sqrt{\text{cosh}r}}\sum_{n=0}^{\infty}(\text{tanh}r)^{n}\frac{\sqrt{(2n)!}}{2^{n}n!}\vert2n\rangle,\label{eq:smss}
\end{equation}
where $\vert n\rangle$ denotes the number state of the considered
harmonic oscillator. The state in Eq.~(\ref{eq:smss}) has enhanced
position or momentum delocalization is given by 
\begin{equation}
\Delta x=\delta xe^{-r},\qquad\Delta p=\delta pe^{r},
\end{equation}
respectively (with $\delta x$ and $\delta p$ denoting the zero-point-motions defined below Eq.~\eqref{eq:modepxpy}). Such states are
not energy eigenstates, and hence, we have to take into account their
time-evolution in the harmonic traps. However, as the initial state in Eq.~\eqref{eq:squeezed} is Gaussian, and
the interaction in Eq.~\eqref{eq:quadratic} is quadratic in the
operators, the state will remain Gaussian also at any later time.

The Heisenberg equations of motion for the modes of the harmonic oscillators
evolve as~\citep{canosa2015exact}:

\begin{alignat}{1}
\hat{a}(t) & =c_{0}(t)\hat{a}+c_{+}(t)\hat{b}+c_{-}(t)\hat{b}^{\dagger},\label{eq:at}\\
\hat{b}(t) & =c_{0}(t)\hat{b}-c_{+}(t)\hat{a}-c_{-}(t)\hat{a}^{\dagger},
\end{alignat}
where $\hat{a}\equiv\hat{a}(0)$, $\hat{b}\equiv\hat{b}(0)$. The
time-dependent coefficients are given by:~\citep{canosa2015exact}

\begin{alignat}{1}
c_{0}(t) & =\text{cos}(\omega_{\text{e}}t)-i\frac{\omega_{\text{m}}}{\omega_{\text{e}}}\text{sin}(\omega_{\text{e}}t),\\
c_{\pm}(t) & =g_{\pm}\frac{\omega_{\text{m}}}{\omega_{\text{e}}}\text{sin}(\omega_{\text{e}}t),
\end{alignat}
where we have defined the effective frequency $\omega_{\text{e}}=\sqrt{\omega_{\text{m}}^{2}+g_{+}^{2}-g_{-}^{2}}$.
We can now readily compute the time dependency of the entanglement
entropy. Inserting Eq.~\eqref{eq:at} in Eq.\eqref{eq:ft} we find that
the problem reduces to evaluating the expectation values of the initial state in Eq.~\eqref{eq:squeezed}. In particular, to complete the analysis, we use the following expectation values~\cite{gerry2023introductory,leonhardt2010essential}:
\begin{alignat}{1}
\langle\hat{a}\hat{a}\rangle & =\langle\hat{b}\hat{b}\rangle=\text{-sinh}(r)\text{cosh}(r),\label{eq:aa}\\
\langle\hat{a}{}^{\dagger}\hat{a}{}^{\dagger}\rangle & =\langle\hat{b}^{\dagger}\hat{b}^{\dagger}\rangle=\text{-sinh}(r)\text{cosh}(r),\label{eq:adad}\\
\langle\hat{a}\hat{a}{}^{\dagger}\rangle & =\langle\hat{b}\hat{b}^{\dagger}\rangle=\text{cosh}^{2}(r),\label{eq:aad}\\
\langle\hat{a}{}^{\dagger}\hat{a}\rangle & =\langle\hat{b}^{\dagger}\hat{b}\rangle=\text{sinh}^{2}(r),\label{eq:ada}
\end{alignat}
where the expectation values are computed with respect to the SMSV state in Eq.~\eqref{eq:smss}.

Inserting Eq.~\eqref{eq:ft} in Eq.~\eqref{eq:st}, and neglecting higher
order terms in $g_{x,p}/\omega_{\text{m}}$, we eventually find the
formula for the time-dependent entanglement entropy $S(t)$:

\begin{alignat}{1}
S(t)\approx & -\frac{A(t)}{2\text{\ensuremath{\omega_{\text{m}}^{2}}}}\sin^{2}(\text{\ensuremath{\omega_{\text{m}}}}t)\left(\text{ln}\left[\frac{A(t)}{2\omega_{\text{m}}^{2}}\sin^{2}(\text{\ensuremath{\omega_{\text{m}}}}t)\right]-1\right).\label{eq:Stbig}
\end{alignat}
where
\begin{alignat}{1}
A(t)= & g_{p}^{2}-4g_{p}\text{\ensuremath{g_{x}}}+g_{x}^{2}+\left(g_{p}^{2}+g_{x}^{2}\right)\cos(2\text{\ensuremath{\omega_{\text{m}}}}t)\nonumber \\
 & +2\left(g_{p}^{2}e^{4r}+g_{x}^{2}e^{-4r}\right)\sin^{2}(\text{\ensuremath{\omega_{\text{m}}}}t).\label{eq:At}
\end{alignat}

We analyze the temporal behaviour of Eq.~\eqref{eq:Stbig} in Fig.~\ref{fig:timecaseAppendix}.
We first find that the behaviour remains qualitatively similar as we
change the frequency $\omega_{\text{m}}$. The entanglement entropy
$S(t)$ has the maximum at $\omega_{\text{m}}t=\pi/2$ in all cases.
Setting $\omega_{\text{m}}t=\pi/2$ we find that the maximum entanglement
increases both for $r<0$ (i.e., spatial delocalization $\Delta x>\delta x$)
as well as for $r>0$ (momentum delocalization $\Delta p>\delta p$).
This is not surprising as a squeezed state breaks the symmetry
of the ground state, leaving it more exposed to TMS with either positive
or negative values.

The generated entanglement becomes negligible when the condition $A = 0$ is met (at $t=\pi/(2\omega_\text{m})$). In this case we are in an entanglement dip, such that $S(t)= 0\,\,\forall t$, as can be noted by computing the limit of Eq.~\eqref{eq:Stbig}:
\begin{equation}
    \lim_{A\rightarrow 0} S(t)=0.
\end{equation}
In particular, if we set $t=\pi/(2\omega_\text{m})$ in Eq.~\eqref{eq:At}, and impose $A=0$, we find the simple condition for the location of the dip:
\begin{equation}
g_x e^{-2r}=g_p e^{2r}.\label{eq:scondition}
\end{equation}
If we set $r=0$ in Eq.~\eqref{eq:scondition} we recover the condition $g_x = g_p$ emerging at the level of the Hamiltonian in Eqs.~\eqref{eq:HamiltonianModes}-\eqref{eq:couplings} as highlighted in the main text.
In other words, the squeezing effectively changes the quadratic position and momentum couplings resulting in the modified condition for the dip. 
We observe the shifted dip location according to Eq.~\eqref{eq:scondition} in Fig.~\ref{fig:timecase}.

To summarize, the 0PN position coupling
$\hat{x}_{\text{A}}\hat{x}_{\text{B}}$ and the 1PN momentum coupling
$\hat{p}_{\text{A}}\hat{p}_{\text{B}}$ induce two-mode squeezing
(TMS) with an opposite sign of the squeezing parameter $\xi$. The
0PN and 1PN coupling are a source of TMS with squeezing parameter
$\xi=+1$ and $\xi=-1$, which would individually generate TMS entangled
states. However, when the two effects combine, they cancel, resulting
in a strong suppression of gravitationally induced entanglement.

As a safety check, we consider the case where the
position and momentum coupling match such that $g_{-}=g_{x}-g_{p}$
vanishes, and we would thus expect a strong suppression of entanglement
generation. We define 
\begin{equation}
\omega_{0}\equiv c/(\sqrt{2}d),\qquad g_{0}\equiv\sqrt{2}Gm/(c^{2}d),
\end{equation}
and find that when $\omega_{\text{m}}=\omega_{0}$ we have equal couplings
$g_{x}=g_{p}=g_{0}$. Using this symmetric coupling regime, we find
from Eq.~(\ref{eq:At}) a simplified expression.

\begin{equation}
A(t)\approx8g_{0}^{2}\sinh^{2}(2r)\sin^{2}(\text{\ensuremath{\omega_{\text{m}}}}t).\label{eq:Aapprox}
\end{equation}
Setting $\omega_{\text{m}}t=\pi/2$ and taking the limit $r\rightarrow0$
in Eq.~(\ref{eq:Stbig}) with $A(t)$ from Eq.~(\ref{eq:Aapprox})
we recover that the entanglement entropy vanishes (case corresponding
to the dip found in Fig.~\ref{fig:simple1} for the ground state).

\end{document}